%% file: main.tex
\documentclass[sigconf]{acmart}
\AtBeginDocument{%
  \providecommand\BibTeX{{%
    \normalfont B\kern-0.5em{\scshape i\kern-0.25em b}\kern-0.8em\TeX}}}

\copyrightyear{2024}
\acmYear{2024}
\setcopyright{rightsretained}
\acmConference[ICSE-NIER'24]{New Ideas and Emerging Results }{April 14--20, 2024}{Lisbon, Portugal}
\acmBooktitle{New Ideas and Emerging Results (ICSE-NIER'24), April 14--20, 2024, Lisbon, Portugal}\acmDOI{10.1145/3639476.3639768}
\acmISBN{979-8-4007-0500-7/24/04}

\usepackage[normalem]{ulem}  
\useunder{\uline}{\ul}{}  
\usepackage{booktabs} 
\usepackage{multirow} 

\input{utils/macros}

\begin{document}

\title{
Which Syntactic Capabilities Are Statistically Learned by \\ Masked Language Models for Code?
}

\author{Alejandro Velasco, David N.~Palacio, Daniel Rodriguez-Cardenas and Denys Poshyvanyk}
\email{{svelascodimate, danaderpalacio, dhrodriguezcar, dposhyvanyk}@wm.edu}
\orcid{0000-0002-4829-1017}
\affiliation{%
  \institution{William \& Mary}
  \city{Williamsburg}
  \state{Virginia}
  \country{USA}
}

\renewcommand{\shortauthors}{Velasco, et al.}

\begin{abstract}

\input{text/0_abstract}
\end{abstract}



\maketitle

\input{text/1_introduction}

\input{text/2_background}

\input{text/3_design}

\input{text/4_use_case}
\input{text/5_results}
\input{text/7_conclusions}

\input{text/8_ACK}


\bibliographystyle{ACM-Reference-Format}
\bibliography{main}

\end{document}

%% file: utils/macros.tex
\usepackage{graphicx}
\usepackage{textcomp}
\usepackage[most]{tcolorbox} 
\def\BibTeX{{\rm B\kern-.05em{\sc i\kern-.025em b}\kern-.08em
    T\kern-.1667em\lower.7ex\hbox{E}\kern-.125emX}}

\usepackage[most]{tcolorbox} 
\usepackage{xspace} 
\usepackage[normalem]{ulem}  
\useunder{\uline}{\ul}{}  
\usepackage{booktabs} 
\usepackage{multirow} 
\usepackage{subcaption} 
\usepackage[noend]{algpseudocode}
\usepackage{enumitem}
\usepackage{adjustbox}
\usepackage{wrapfig}
\usepackage[ampersand]{easylist}
\usepackage[most]{tcolorbox}

\usepackage{colortbl}

\definecolor{main}{HTML}{5989cf}    
\definecolor{sub}{HTML}{cde4ff}     

\newtcolorbox{boxB}{
    fontupper = \bf\color{main}\footnotesize, 
    boxrule = 0.5pt,
    colframe = main,
    rounded corners,
    arc = 5pt   
}

\newtcolorbox{boxD}{
    fontupper = \small, 
    colback = sub, 
    colframe = main, 
    boxrule = 0pt, 
    toprule = 2pt, 
    bottomrule = 2pt 
}

\newtcolorbox{boxH}{
    fontupper = \small, 
    colback = sub, 
    colframe = main, 
    boxrule = 0pt, 
    leftrule = 6pt 
}

\newtcolorbox{boxG}{
    enhanced,
    boxrule = 0pt,
    colback = sub,
    borderline west = {1pt}{0pt}{main}, 
    borderline west = {0.75pt}{2pt}{main}, 
    borderline east = {1pt}{0pt}{main}, 
    borderline east = {0.75pt}{2pt}{main}
}    

\newtcolorbox{boxK}{
    fontupper = \small,
    sharpish corners, 
    boxrule = 0pt,
    toprule = 1.0pt, 
    enhanced,
    fuzzy shadow = {0pt}{-2pt}{-0.5pt}{0.5pt}{black!35} 
}

\definecolor{MidnightBlue}{HTML}{006895}

\newcommand*\circled[1]{\tikz[baseline=(char.base)]{
            \node[shape=circle,draw,inner sep=0.5pt] (char) {#1};}}

\newboolean{showcomments}

\setboolean{showcomments}{true}

\ifthenelse{\boolean{showcomments}}
  {\newcommand{\nb}[2]{
    \fbox{\bfseries\sffamily\scriptsize#1}
    {\sf\small$\blacktriangleright$\textit{#2}$\blacktriangleleft$}
   }
   
  }
  {\newcommand{\nb}[2]{}
   
  }

\newcommand{\ie}{\textit{i.e.,}\xspace}
\newcommand{\eg}{\textit{e.g.,}\xspace}

\newcommand{\etal}{et al.\xspace}

\newcommand{\aka}{\textit{a.k.a.}\xspace}	

\newcommand{\syntaxeval}{Syntax\textit{Eval}\xspace}

\newcommand{\llms}{LLMs\xspace}

\newcommand{\mlm}{MLM\xspace}
\newcommand{\mlms}{MLMs\xspace}

\newcommand{\cfg}{CFG\xspace}
\newcommand{\asts}{ASTs\xspace}

\newcommand{\scm}{SCM\xspace}

\newcommand{\iqr}{$iqr$\xspace}
\newcommand{\std}{$std$\xspace}

\newcommand{\docode}{$do_{code}$\xspace}



%% file: text/0_abstract.tex
This paper discusses the limitations of evaluating Masked Language Models (\mlms) in code completion tasks. We highlight that relying on accuracy-based measurements may lead to an overestimation of models' capabilities by neglecting the syntax rules of programming languages. To address these issues, we introduce a technique called \syntaxeval in which \textit{Syntactic Capabilities} are used to enhance the evaluation of \mlms. \syntaxeval automates the process of masking elements in the model input based on their Abstract Syntax Trees (ASTs). We conducted a case study on two popular \mlms using data from GitHub repositories. Our results showed negative causal effects between the node types and \mlms' accuracy. We conclude that \mlms under study fail to predict some syntactic capabilities. 

%% file: text/1_introduction.tex
\section{Introduction}\label{sec:introduction}

Large language models have illustrated convincing performance across a range of different software engineering (SE) tasks~\cite{Tufano.MSR.2018, SANER.2019,White2016clones,Watson:ICSE20,ciniselli2021empirical,Mastropaolo2021StudyingTasks,Tufano:icse2021,Chen2019sequencer}. In particular, \textit{code generation} has been an important area of research for SE tasks such as code completion~\cite{MSR-Completion}.
\textit{Code completion} is a disciplined technique for generating missing \textit{syntactic features} of an incomplete snippet based on its semantic and structural context \cite{bruch_learning_2009}. These syntactic features usually adopt the form of identifiers, function names, conditionals, or parameters depending on the granularity of the snippet. Software researchers are particularly interested in improving code completion to optimize time spent during the development and maintenance cycles \cite{han_code_2009,han_code_2011}. Numerous studies have investigated code completion automation using machine learning \cite{bruch_learning_2009,white_toward_2015,raychev_code_2014,6227135}. Current research has focused on exploiting deep learning representations using LSTMs \cite{svyatkovskiy_pythia_2019}, GPT \cite{svyatkovskiy_intellicode_2020}, RoBERTa \cite{liu2019roberta}, and T5 \cite{clement_pymt5_2020,ciniselli_empirical_2021}.


Masked Language Models (\mlms) have been recently used for code completion tasks demonstrating promising results (an avg. accuracy of $38.7\%$ in perfect predictions) at different masking levels (\ie Token, Construct, and Block) \cite{ciniselli_empirical_2021}. Some studies suggest that \mlms statistically learn the underlying structure of Abstract Syntax Trees (\asts) at certain degree \cite{mohammadkhani_explainable_nodate, karmakar_inspect_2023, wan_what_2022}. Yet, given the high accuracy achieved by \mlms \cite{ciniselli_empirical_2021}, few attempts have been made to investigate the role of \textbf{Syntactic Capabilities} for evaluating code completion. Syntactic Capabilities are \textit{interpretable} prediction estimates for a terminal ($N$) and non-terminal ($\Sigma$) nodes of \asts that are ruled by a \textit{Context Free Grammar} (\cfg) of Programming Languages (PLs) \cite{srimani_nasir_2007}. 

To date, the primary focus on evaluating \mlms has been on the role of \textit{accuracy} as the principal metric, which may lead to erroneous and/or incomplete interpretation of the \textit{syntactic features} embedded in neural architectures \cite{palacio_toward_2023,rabin_towards_2020,wan_what_2022}. Relatively little is understood about incorporating these interpretable prediction estimates into the evaluation of \mlms, hence \textit{current evaluation methods do not help practitioners to decide} whether \mlms are confidently generating code at AST node granularity and to what extent these syntactic features affect general prediction performance. That is, these methods do not reveal information about syntactic capabilities and their causal effects on the overall  \mlms performance. 





Our study attempts to establish the causal connection between syntactic features in the form of AST node types and \mlms' performance. Under this premise, we introduce \syntaxeval, an approach that leverages syntactic capabilities to evaluate how good \mlms infer $N$ and $\Sigma$ AST nodes of a given PL. When evaluating the performance of an \mlm, \syntaxeval selectively masks tokens according to the AST Node types defined by the \cfg. Subsequently, an \mlm predicts the masked tokens. Finally, \syntaxeval measures the \textbf{causal effect} of AST node types on code completion performance. 

Our results suggest that although \mlms are homogeneously predicting individual AST node types with high accuracy, we observed no evidence of effects from syntactic features on \mlms' prediction after controlling for confounding factors. Hence, no causal evidence supports the fact that \mlms are statistically learning syntactic structures with acceptable confidence, contradicting recent studies in the explainability field \cite{wan_what_2022,mohammadkhani_explainable_nodate}. We hope that the results of our work will shed more light on the syntactic capabilities of current \mlms to enable a more systematic and rigorous evaluation of code completion tasks. The contributions of this paper are as follows: 1) a technique for evaluating the extent to which \mlms predict AST structures; 2) a case study that leverages causal analysis to understand how different AST node types influence code completion; 3) experimental data, curated datasets, source code, and complementary statistical analysis used in this research are published in an open-source repository \cite{wm-semerusyntaxeval_2023}.


%% file: text/2_background.tex
\section{ Background \& Related Work}
\label{sec:background}

The accurate identification and generation of code tokens is a widely studied field at the intersection of SE and DL \cite{watson_systematic_2022}. State-of-the-art code generators estimate the token prediction using probabilistic distribution (\ie a Large Language Model (LLMs)) obtained by training on large amounts of code corpora. Put simply, code completion models should statistically approximate the production rules defined by the \cfg. These production rules are recursively applied to terminal $N$ and non-terminal $\Sigma$ nodes to formally define the structure of a PL. For instance, recent explainability studies have claimed that the syntactic structures of code are encapsulated in the internal layers of \llms across software tasks, implying a foundational statistical comprehension of code semantics \cite{hernandez_lopez_ast-probe_2022, wan_what_2022}. In this section, we introduce the concept of \mlms and their current evaluation methods.

\textbf{Masked Language Models for Code.} Considerable research attention has been directed toward the usage of Bidirectional Encoder Representation from Transformers (BERT) on code completion as an attempt to push the predictability boundaries beyond the next token prediction. BERT allows higher granularity syntax structures (\ie entire code statement) to be generated using self-attention layers trained to restore a masked subset of tokens in the input \cite{devlin_bert_2019,ciniselli_empirical_2021}. This peculiar form of training the architecture is known as \textit{denoising autoencoding}, or Masked Language Models (\mlm), which we formalize as $  MLM(C) =  \mathbb{E}_{s\in S} \mathbb{E}_{M\subset s} \left[ \sum_{s_j \in M}\log{p(s_j|\tilde{s})} \right], |M|=|m(s)|$, where a masking rate $m$ (usually $15\%$) is applied on the original sequence $s$ of a training corpus $S$. The model attempts to predict the set of masked tokens $M$ given the corrupted context $\tilde{s}$ (the masked version of $s$) \cite{kaneko2020encoderdecoder}. \mlms for code completion are mostly evaluated using  metrics such as CodeBleu, EM, F1, and Pass@k \cite{hou_large_2023}.  

\textbf{Syntax-Based Evaluation of \mlms.} Due to the unpredictable behavior of \mlms while generating tokens, explainability techniques are complementary evaluative methods for understanding the decision-making process by reducing the uncertainty of the models. Such uncertainty can be controlled by exploring the inner layers of the neural net or performing guided perturbations on models' input \cite{DBLP:journals/corr/abs-2009-11698}. Recent studies have explored the use of structural information as an interpretability tool for pre-trained models for code \cite{palacio_toward_2023}. For instance, Wan \etal \cite{wan_what_2022} conducted an explainability analysis focusing on three aspects: 1) how the self-attention weights align with the syntax structure, 2) whether the syntax structure is encoded in the hidden layers, and 3) how pre-trained models induce syntax structures. Similarly, Mohammadkhani \etal \cite{mohammadkhani_explainable_nodate} propose an eXplainable AI method (attention mechanism) on three downstream code tasks: 1)  code generation, 2) refinement, and 3) translation. Previous findings imply that Encoder-based models can effectively extract detailed syntactic information using self-attention mechanisms. We used prior observations about encoded information of ASTs to formulate an evaluative approach based on measuring the prediction performance of syntactic capabilities directly from (non)terminal nodes.

%% file: text/3_design.tex
\section{Syntactic Causal Evaluation}
\label{sec:design}
\syntaxeval is an evaluative approach organized into two distinct parts. The first part estimates a fine-grained performance, grouped by AST node types, for a given \mlm (\ref{rq:performance}). The second part adopts causal interpretability theory to quantify the influence of previously estimated AST node types on the accuracy of the model (\ref{rq:causal}). 

\begin{figure}[t]
		\centering
  \vspace{-1.1em}
  \includegraphics[width=0.46\textwidth]{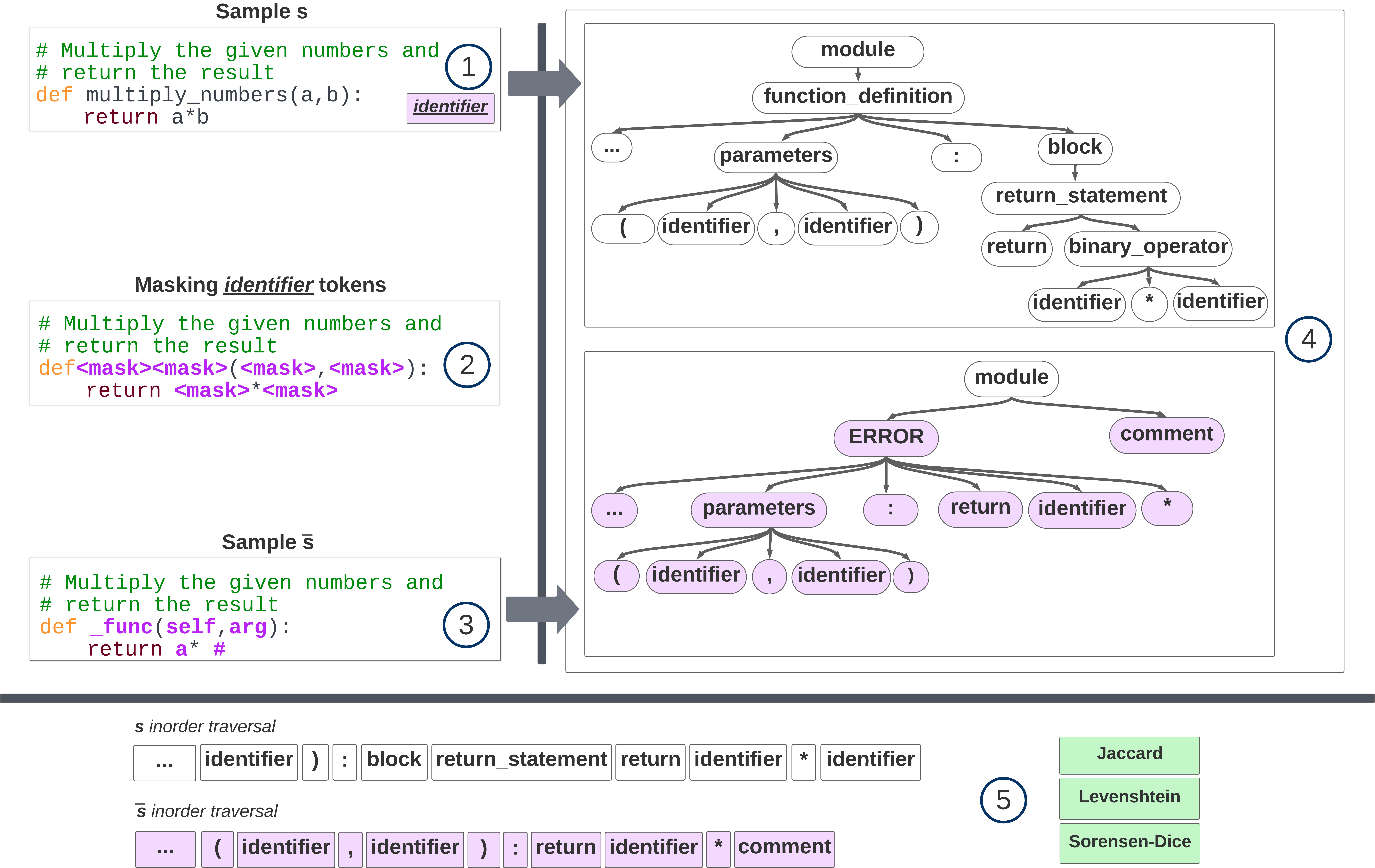}
		\caption{\syntaxeval Process for the \textit{\small identifier} AST Node.}
        \label{fig:pipeline}
        \vspace{-1.5em}
\end{figure}

\textbf{Evaluating Syntactic Capabilities.} Fig.~\ref{fig:pipeline} depicts the process of evaluating syntactic capabilities for code completion using \mlms. This evaluative process is comprised of five steps. Firstly, we must define a set of AST node types $C$ to be analyzed. This set of AST node types is ruled by Python \cfg adopting the form of $C = N \cup \Sigma$. Then we search for the positions of these node types $C$ in the code sequence after iterating for each sample $s_i$ (\ie snippet) of a given ground truth $S$ (Fig.~\ref{fig:pipeline}- \circled{1}). Secondly, detected tokens, which correspond to the previously defined set $C$, are masked with the label \textbf{$<$mask$>$} (Fig.~\ref{fig:pipeline}- \circled{2}). Thirdly, we use an \mlm to infer the masked tokens for each sample $s_i$ obtaining a set $\Bar{S}$ of predicted samples $\Bar{s_i}$. Fourthly, we parse the AST of $s_i$ and $\Bar{s_i}$ to generate a list of extracted nodes for the ground truth and predicted samples using the in-order traversal algorithm (Fig.~\ref{fig:pipeline}- \circled{4}). Finally, we compare both ground truth $s_i$ and predicted $\Bar{s_i}$ lists of extracted nodes by computing three similarity metrics for each sample (\ie \textbf{Jaccard}, \textbf{Levenshtein} \& \textbf{Sorensen-Dice}) (Fig.~\ref{fig:pipeline}- \circled{5}).

\textbf{Computing Causal Interpretability.} Causal Inference has been adopted to complement the assessment of \llms by controlling for confounding factors in code data. Palacio \etal \cite{palacio_toward_2023} introduce \docode, a post hoc interpretability methodology that explains model predictions by providing causal explanations. These explanations are generated by estimating the effect of binary interventions $T$, such as masking random tokens $T_0$ versus masking AST node types $T_1$, on \mlms' performance. Specifically, in \syntaxeval, the treatment $T_1$ refers to samples that are masked on AST node tokens $C$, while the control $T_0$ refers to samples that are randomly masked on any position. The control $T_0$ preserves the same number of masked tokens as in $T_1$.

\syntaxeval formulates a \textbf{Structural Causal Model} (\scm), which is a graphical model composed of outcomes, treatments, and confounders \cite{Pearl2009Causality}, to explain a set of \textit{potential outcomes} $Y$ (\eg Jaccard, Levenshtein, Sorensen-Dice) in terms of treatments $T$ (\ie masked AST node types) by controlling for a set of code confounders $Z$ to avoid \textit{spurious correlations}. These code confounders consist of seven variables, which include the \# of parsing errors, the height of the AST, the \# of nodes, the \# of whitespaces, the \# of lines of codes, the cyclo complexity, and the token counts. Finally, \syntaxeval computes the \textit{Average Treatment Effect} ($\tau$) of a treatment $T$ has on the outcomes $Y$ after controlling for confounders $Z$. In other words, we want to estimate the expected value $\tau = \mathbf{E}[\tau(Z)] = \mathbf{E}[Y|do(T_1)] - \mathbf{E}[Y|do(T_0)] = \mathbf{E}[Y_1 - Y_0]$. The variables $Y_1, Y_0$ refer to potential outcomes observed under the treatments $T_1,T_0$. For the sake of brevity, we do not discuss the details of treatment effects computations. However, these effects are approximated using \textit{propensity score methods} after applying the \textit{the back-door criterion} \cite{Sharma2021DoWhyAssumptions}.

%% file: text/4_use_case.tex
\section{Case Study Design}
\label{sec:approach}

This section outlines the methodology employed to consider the potential influence of syntactic capabilities on the evaluation of \mlms, we conducted a case study on two popular architectures to explore the following RQs:

\begin{enumerate}[label=\textbf{RQ$_{\arabic*}$}, ref=\textbf{RQ$_{\arabic*}$}, wide, labelindent=5pt]\setlength{\itemsep}{0.2em}
      \item \label{rq:performance} {\textbf{[Performance]} 
      \textit{How good are \mlms at predicting AST nodes?}} 
      \item \label{rq:causal} {\textbf{[Causality]} 
      \textit{How do node types impact \mlms' performance?}}
\end{enumerate}

\textbf{Data Collection:} To mitigate the risk of data snooping, we curated our testbed with $~50k$ Python snippets. This testbed exclusively comprises commits executed between January 01, 2022 and January 01, 2023. We collected the snippets from newly added or updated Python Github repositories with over 1k stars scoring. Additionally, we discarded duplicated samples by referring to the history of the commits. The testbed also contains complementary code features (\eg LoC, CYCLO, and \# of nodes), these features were extracted using \textit{Galeras} pipeline \cite{rodriguez-cardenas_benchmarking_2023}. \textbf{Masked Language Models:} We evaluated two encoder-based transformers trained on \textit{CodeSearchNet} \cite{husain_codesearchnet_2019} with different hyperparameters (see Tab.~\ref{tab:models}). These encoders have been assessed in prior studies in which they were found to capture structural information  \cite{wan_what_2022}, \cite{wan_what_2022}, and \cite{mohammadkhani_explainable_nodate}. \textbf{Node Types:} Tree-sitter \cfg defines $196$ AST node types $C$ for Python. For the sake of simplicity, we selected a subset of terminal and non-terminal nodes defined in Python's \cfg as depicted in the first column of Tab.~\ref{tab:correlations}. The subset entails the most basic syntactic structures for control, iteration, operators, and functional programming. This study showcases the nodes that exhibited the most interesting behavior. We chose Python for code completion experiments due to its extended use in recent studies.

\input{tabs/use_case_1_models} 

\textbf{Evaluation Methodology.} To address \ref{rq:performance}, we estimated syntactic capabilities of $M_1$ and $M_2$ encoders using $~8K$ randomly selected samples from the collected testbed. \syntaxeval masks the associated tokens for each chosen node type ($T_1$) and subsequently uses the \mlm to infer the missing elements. Then, we compute normalized similarity distances (\ie Jaccard, Levenshtain, and Sorence-Dice) between the AST in-order traversal of both the predictions and the ground truth. Global results indicate the average prediction accuracy (\ie normalized distance) for all node types within $C$. In contrast, local results detail the prediction accuracy for individual node types.

To address \ref{rq:causal}, \syntaxeval computes the \textit{Average Causal Effect} between syntactic capabilities and \mlms' performance. This method consists of estimating $\tau$ using treatments $T_1$ and $T_0$ (\ie tokens randomly masked) while controlling for confounders in $Z$ (code features in Data Collection), to mitigate the presence of \textit{spurious correlations}. The removal of confounding bias can be formally achieved using both an \scm and the $do$-operator introduced by Pearl \etal \cite{Pearl2009Causality}. To verify the robustness of our SCM, we computed \textit{placebo} refutations, which is a method that fakes an unrelated treatment by re-estimating the causal effects. That is, we assessed that the causal effects of the fake treatment on the outcome were close to zero. Moreover, to ensure a balanced distribution of randomly masking tokens within $T_0$, we created 20 distinct variations for each sample. Afterward, we computed the average of the resulting similarity scores. Finally, to ensure statistical significance, we \textit{bootstrapped} the similarity scores using the $mean$ for $500$ samples per node type.

%% file: tabs/use_case_1_models.tex
\begin{table}[]

\centering
\caption{Evaluated Encoder-Based Transformers.}
\label{tab:models}
\vspace{-0.5em}
\scalebox{0.9}{

\setlength{\tabcolsep}{4pt} 

\begin{tabular}{lllll}
\hline
\textbf{Id} & \textbf{\mlm}     & \textbf{Size} & \textbf{Layers} & \textbf{Vocab.} \\ \hline
$M_1$          & CodeBERTa-small-v1 \cite{liu_roberta_2019}  & 84M           & 6                      & 52,000                    \\
$M_2$          & codebert-base-mlm \cite{feng2020codebert} & 125M          & 12                     & 50,265                    \\ \hline
\end{tabular}

} 
\vspace{-1em}
\end{table}

%% file: text/5_results.tex
\section{Results \& Discussion}
\label{sec:results}

The aim of this study is to determine the effect of Syntactic Capabilities, in the form of interpretable prediction estimates for node types, on the prediction performance of \mlms. We concentrated on evaluating Encoder-based Transformers beyond accuracy.

\subsection{\ref{rq:performance} Syntactic Capabilities Performance}

\textbf{Global Results.} A cursory glance at Tab.~\ref{tab:correlations} reveals that control groups $T_0$ of each performance metric are \textit{not significantly different} from treatments $T_1$ for both encoders. For example, the control median values greater than $0.8$ are within the interquartile range (\iqr) $0.78\pm0.22$ of the corresponding treatment. Furthermore, the standard deviation (\std) values of the performance are predominantly more dispersed in the treatments than in the control. For example, the $T_1$ \std of $M_1$ Jaccard is $0.21$, while the $T_0$ is $0.17$. Appealingly, all average values of performance are above $0.5$, this indicates that $M_1$ and $M_2$ models are predicting masking tokens with high confidence despite the group treatments $T$. Although the median global performance has consistently high accuracy among the metrics ($>0.8$), the average separation values between the $T$ groups are not significant with an average median distance of $0.096$ and $0.06$ for $M_1$ and $M_2$ respectively. However, a preliminary analysis for node types estimations suggests that $M_1$ and $M_2$ have a tendency to not statistically learn syntactic-oriented masked tokens $T_1$. Our findings reveal a subtle inclination towards predicting random masked tokens over syntactic-oriented ones.




\begin{figure}[ht]
		\centering
  \vspace{-1em}
  \includegraphics[width=0.4\textwidth]{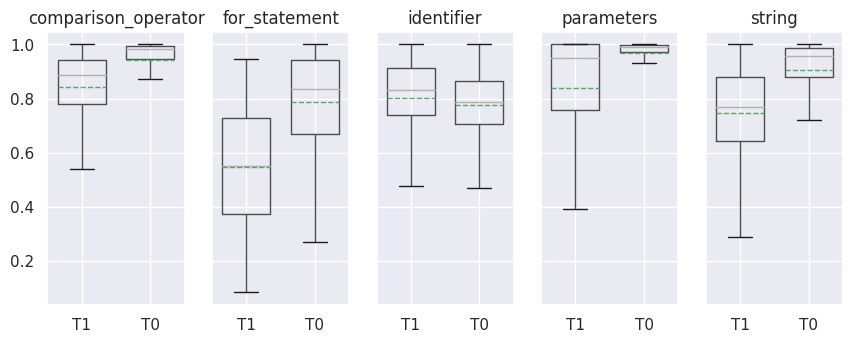}
		\caption{$T_0$ vs. $T_1$ Local Jaccard for \textit{Nodes} using $M_1$.}
        \label{fig:local_results}
        \vspace{-1em}
\end{figure}

\begin{figure}[ht]
		\centering
  \vspace{-1em}
  \includegraphics[width=0.4\textwidth]{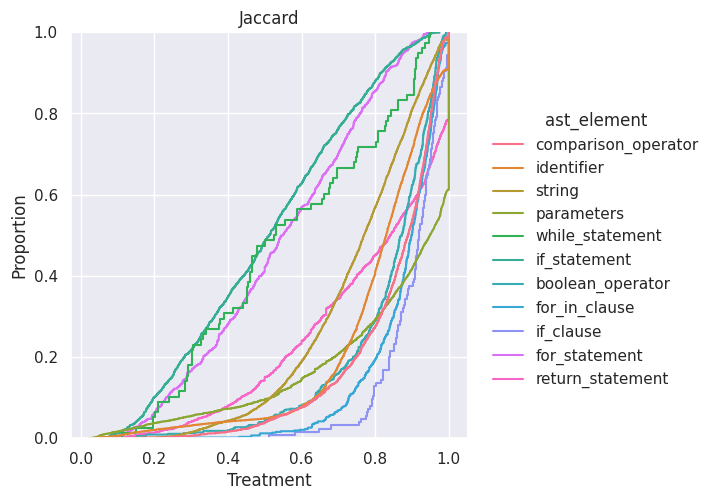}
		\caption{Syntactic Capabilities Statistically Learned by $M_1$.}
    \label{fig:ecdf}
    \vspace{-1em}
\end{figure}

\textbf{Local Results.} Fig.~\ref{fig:local_results} shows the Jaccard performance statistical behavior across some selected node types for $M_1$. Due to the non-overlapping \iqr between the $T_0$ and $T_1$, we observed a significant difference between treatment groups in the performance distribution for the nodes \textit{\small comparison\_operator} and \textit{\small string}{, revealing that $M_1$ struggles at predicting tokens associated with such types in contrast to random masked tokens}. We found that \textit{\small identifier} was the only node type that performed better in the treatment than the control group. Fig.~\ref{fig:ecdf} presents the Empirical Cumulative Distribution (ECD) plots of $T_1$ Jaccard distance across selected node types. We observed that \textit{\small if\_clause} was remarkably achieved with the highest score prediction ($0.9$) at the lowest percentage of the population ($42\%$ of the samples in the testbed). Conversely, \textit{\small for\_statement} was the most difficult node to predict across the population. We believe that \mlms struggle to predict these previous nodes due to their complexity. A node is complex when its block has incorporated other node types. 

\begin{boxK}
\textit{\ref{rq:performance}:} 
\mlms tend to complete missing AST-masked tokens with acceptable accuracy ($>0.5$). However, the reported performance suffers from high variability ($\pm 0.21$) making the prediction process less confident compared to completing randomly masking tokens.
\end{boxK}

\subsection{\ref{rq:causal} Causal Evaluation Effect}

\input{tabs/results_causal_effects} 
This study used a quantitative causal technique to analyze the influence of masking binary treatments (\ie AST and random) on the performance of both $M_1$ and $M2$ transformers after defining the Structural Causal Model of the problem. To draw a causal link between syntactic features (\ie AST nodes) and performance metrics (\ie Jaccard, Levenshtain, and SD), we expect to observe a \textbf{positive causal effect}. A positive effect would indicate that syntactic features $C$ are affecting models' performance and AST nodes would be statistically learned by \mlms. On the other hand, a \textbf{negative causal effect} would imply that randomly masked tokens have more influence on the performance. That is, tokens without any particular syntactic order are being predicted accurately.    

Unlike previous assumptions, it can be inferred from Tab.~\ref{tab:correlations} that the control group (\ie masking random treatment) is having more impact on \mlms' performance than the actual syntactic features. For instance, a set of samples masked for \textit{\small for\_statement} tokens are underperforming (\aka negative effects) compared to the same set but randomly masked tokens. This suggests that although transformers are predicting AST node types with confidence (see Fig.~\ref{fig:ecdf}, these syntactic features are not particularly relevant compared to predicting any other set of unstructured tokens in the snippet (see gray areas in Tab.~\ref{tab:correlations}). These findings tend to corroborate Karmakar \etal research \cite{karmakar_inspect_2023} in which \mlms do not fully grasp the syntax and structural aspects of code. Our findings offer an alternative perspective compared to claims made by other probing approaches \cite{troshin_probing_2022,ma_are_2023}. For example, Hernandez Lopez \etal \cite{hernandez_lopez_ast-probe_2022} argue for the presence of a syntax subspace within the hidden layers that encode structures of PLs. {Similarly, Toufique \etal \cite{ahmed_towards_2023} outline that pre-trained language models learn robust representations of code semantics, which implies a deep understanding of syntax elements from the source code.}

\begin{boxK}
\textit{\ref{rq:causal}:}
The performance of \mlms is negatively impacted by AST-masked tokens ($\tau<-0.1$). Our causal analysis yielded no signs of Transformers' performance being affected or guided by syntactic features, contradicting SOTA explainability findings.
\end{boxK}

%% file: tabs/results_causal_effects.tex
\begin{table}[]
\centering
\caption{Global Perf. and Causal Effects for $M_1$ and $M_2$.}
\vspace{-0.5em}
\label{tab:correlations}

\scalebox{0.6}{

\setlength{\tabcolsep}{4pt} 
\begin{tabular}{clcccccc}
\hline
\textbf{Performance} &  & \multicolumn{2}{c}{\textbf{Jaccard}} & \multicolumn{2}{c}{\textit{\textbf{Levenshtein}}} & \multicolumn{2}{c}{\textit{\textbf{Sorensen-Dice}}} \\ \hline
\mlms &
   &
  $M_1$ &
  $M_2$ &
  $M_1$ &
  $M_2$ &
  $M_1$ &
  $M_2$ \\ \hline
\textit{\textbf{Treatments}} &
   &
  \multicolumn{6}{c}{\textit{\textbf{Performance Metric $Y$   [avg $\pm$ std]*}}} \\ \hline
$T_0$ &
   &
  0.88 $\pm$ 0.17 &
  0.84 $\pm$ 0.16 &
  0.87 $\pm$ 0.16 &
  0.83 $\pm$ 0.17 &
  0.92 $\pm$ 0.1 &
  0.89 $\pm$ 0.12 \\
$T_1$ &
   &
  0.78 $\pm$ 0.21 &
  0.76 $\pm$ 0.21 &
  0.78 $\pm$ 0.22 &
  0.76 $\pm$ 0.21 &
  0.85 $\pm$ 0.17 &
  0.84 $\pm$ 0.17 \\ \hline
\textit{\textbf{AST Node Type $C$}} &
   &
  \multicolumn{6}{c}{\textit{\textbf{CausalEffect $\tau$}}} \\ \hline
\textit{boolean\_operator} &
   &
  -0.083 &
  \cellcolor[HTML]{EFEFEF}-0.150 &
  -0.069 &
  -0.136 &
  -0.048 &
  -0.095 \\
\textit{comparison\_operator} &
   &
  \cellcolor[HTML]{EFEFEF}-0.186 &
  -0.027 &
  -0.179 &
  -0.018 &
  -0.126 &
  -0.015 \\
\textit{for\_in\_clause} &
   &
  \cellcolor[HTML]{EFEFEF}-0.059 &
  -0.053 &
  -0.050 &
  -0.045 &
  -0.034 &
  -0.029 \\
\textit{for\_statement} &
   &
  \cellcolor[HTML]{EFEFEF}-0.269 &
  -0.101 &
  -0.193 &
  -0.041 &
  -0.243 &
  -0.083 \\
\textit{identifier} &
   &
  0.016 &
  \cellcolor[HTML]{EFEFEF}-0.075 &
  0.001 &
  -0.073 &
  0.010 &
  -0.039 \\
\textit{if\_clause} &
   &
  \cellcolor[HTML]{EFEFEF}-0.070 &
  -0.040 &
  -0.058 &
  -0.036 &
  -0.037 &
  -0.022 \\
\textit{if\_statement} &
   &
  \cellcolor[HTML]{EFEFEF}-0.163 &
  -0.118 &
  -0.140 &
  -0.093 &
  -0.116 &
  -0.095 \\
\textit{parameters} &
   &
  \cellcolor[HTML]{EFEFEF}-0.140 &
  -0.048 &
  -0.127 &
  -0.046 &
  -0.087 &
  -0.029 \\
\textit{return\_statement} &
   &
  \cellcolor[HTML]{EFEFEF}-0.144 &
  -0.121 &
  -0.118 &
  -0.113 &
  -0.087 &
  -0.075 \\
\textit{string} &
   &
  \cellcolor[HTML]{EFEFEF}-0.156 &
  -0.168 &
  -0.102 &
  -0.145 &
  -0.118 &
  -0.116 \\
\textit{while\_statement} &
   &
  \cellcolor[HTML]{EFEFEF}-0.200 &
  -0.096 &
  -0.139 &
  -0.009 &
  -0.186 &
  -0.077 \\ \hline
\end{tabular}

} 

{\scriptsize * Medians are $>0.8$. The biggest causal effect $\tau$ for each node type is in gray. \par}
\vspace{-0.45cm}
\end{table}

%% file: text/7_conclusions.tex
\section{Conclusion \& Future Plans}
\label{sec:conclusion}


Our negative causal effect results corroborate recent findings that show flaws when claiming that \mlms are \textit{understanding} syntax rules of PLs. Such effects amplify the disparities between Natural and Programming languages, underscoring the need for tailored representations in deep learning architectures. These findings pave the way for future research to evaluate \textit{semantic} capabilities in the form of \textit{recursions, dead code, or code smells}. We also highlight the necessity to delve deeper into understanding why \mlms are more adept at predicting random masked tokens than syntax-based tokens. This tendency may be linked to the models' pre-training objectives, which frequently involve masking random tokens at a certain rate \cite{devlin_bert_2019}.

%% file: text/8_ACK.tex
\section{Acknowledgements}
\label{sec:ack}

This research has been supported in part by the NSF CCF-2311469, CNS-2132281, CCF-2007246, and CCF-1955853. We also acknowledge support from Cisco Systems. Any opinions, findings, and conclusions expressed herein are the authors’ and do not necessarily reflect those of the sponsors.